\documentclass{IEEEtran}
\usepackage[T1]{fontenc}
\usepackage[utf8]{luainputenc}
\usepackage{xcolor}
\usepackage{verbatim}
\usepackage{tikz}
\usepackage{amsmath}
\usepackage{amsthm}
\usepackage{amssymb}
\usepackage{graphicx}
\usepackage[bookmarks=true,bookmarksnumbered=true,bookmarksopen=true,bookmarksopenlevel=1,
 breaklinks=false,pdfborder={0 0 1},backref=false,colorlinks=false]
 {hyperref}
\hypersetup{pdftitle={Your Title},
 pdfauthor={Your Name},
 pdfborderstyle=,pdfpagelayout=OneColumn,pdfnewwindow=true,pdfstartview=XYZ,plainpages=false}

\makeatletter

\providecommand{\tabularnewline}{\\}
\providecolor{lyxadded}{rgb}{0,0,1}
\providecolor{lyxdeleted}{rgb}{1,0,0}
\usetikzlibrary{calc}
\newcommand{\lyxobjectsout}[1]{%
  \bgroup%
  \color{lyxdeleted}%
  \tikz{
    \node[inner sep=0pt,outer sep=0pt](lyxdelobj){#1};
    \draw($(lyxdelobj.south west)+(2em,.5em)$)--($(lyxdelobj.north east)-(2em,.5em)$);
  }
  \egroup%
}

\DeclareRobustCommand{\lyxdisplayobjdeleted}[4][]{%
  \ifx#4\empty\else%
     \texorpdfstring{\leavevmode\\\lyxobjectsout{\parbox{\linewidth}{#4}}}{}%
  \fi%
}
\DeclareRobustCommand{\lyxudisplayobjdeleted}[4][]{%
  \ifx#4\empty\else%
     \texorpdfstring{\leavevmode\\\raisebox{-\belowdisplayshortskip}{%
                \lyxobjectsout{\parbox[b]{\linewidth}{#4}}}}{}%
     \leavevmode\\%
  \fi%
}

\theoremstyle{plain}
\newtheorem{thm}{\protect\theoremname}
\theoremstyle{remark}
\newtheorem{rem}[thm]{\protect\remarkname}

\usepackage{cite}
\DeclareMathOperator{\trans}{^{\mathsf{T}}}
\DeclareMathOperator{\diag}{diag}
\newcommand{\herm}{^{\mathsf{H}}}

\makeatother

\providecommand{\remarkname}{Remark}
\providecommand{\theoremname}{Theorem}

\begin{document}
\title{RIS-Aided Monitoring With Cooperative Jamming: Design and Performance
Analysis}
\author{Shuying Lin, Yulong Zou,~\IEEEmembership{Senior Member,~IEEE,} Zhiyang
Li, Tong Wu, Eduard E. Bahingayi,~and Le-Nam Tran,~\IEEEmembership{Senior~Member,~IEEE}\thanks{This work was supported in part by the National Natural Science Foundation
of China (Grant Nos. 62271268 and 62371252), in part by the Jiangsu
Provincial Key Research and Development Program (Grant No. BE2022800),
in part by the Jiangsu Provincial 333 Talent Project, in part by Taighde
Éireann - Research Ireland under Grant numbers 22/US/3847 and 13/RC/2077\_P2,
and in part by China Scholarship Council (CSC). {\emph{(Corresponding
author: Yulong Zou.)}}} \thanks{S. Lin, Y. Zou, and T. Wu are with the School of Telecommunications
and Information Engineering, Nanjing University of Posts and Telecommunications,
Nanjing, China.} \thanks{Z. Li is with the School of Electronics and Information Engineering,
Harbin Institute of Technology, Harbin, China.} \thanks{E.-E. Bahingayi and L.-N. Tran are with the School of Electrical and
Electronic Engineering, University College Dublin, Ireland.}}
\maketitle
\begin{abstract}
Exploiting the potential of physical-layer signals to monitor malicious
users, we investigate a reconfigurable intelligent surface (RIS) aided
wireless surveillance system. In this system, a monitor not only receives
signal from suspicious transmitter via a RIS-enhanced legitimate surveillance
(LS) link but also simultaneously controls multiple jammers to degrade
the quality of the received suspicious signal. To enhance monitoring
performance, it is crucial to improve both the received signal quality
at the monitor and the effectiveness of cooperative jamming (CJ).\textcolor{teal}{{}
}Given\textcolor{teal}{{} }that the surveillance system is aided by
a single RIS, whose phase shift optimization relies on the channel
state information (CSI) of both the LS and CJ links, we utilize partial
CSI%
{} to alleviate the CSI acquisition burden. Specifically, we propose
two RIS-aided monitoring schemes with optimal jammer selection (OJS),
which are differentiated by the CSI knowledge used for the RIS phase
shift design. The first scheme is called RISLO, which is RIS-aided
monitoring with the CSI of LS link and an optimally selected jammer.
The second scheme is called RISCO, which is RIS-aided monitoring with
the CSI of CJ link and an optimally selected jammer. Closed-form expressions
for the surveillance success probability (SSP) are derived for both
schemes. Furthermore, we consider RIS-aided monitoring schemes with
random jammer selection as benchmarks. We further analyze special
cases where the jammers act like passive monitoring by using minimal
power to avoid being found. Also, the impact of RIS is studied under
an asymptotically large number of RIS elements. Numerical results
demonstrate that the proposed OJS strategy significantly enhances
the RIS-aided monitoring performance compared to non-jammer-selection
RISLR and RISCR schemes. However, this improvement comes at the cost
of CSI knowledge and becomes marginal at high jamming power. In addition,
RISLO outperforms RISCO when the suspicious transmitter operates at
low power or when the number of RIS elements is large.
\end{abstract}


\IEEEpeerreviewmaketitle{}

\section{Introduction\protect}\label{sec:Introduction}

Wireless connectivity has become a cornerstone in our modern society
but it also raises serious concerns over information privacy. Consequently,
numerous research endeavors have been made to enhance wireless security
\cite{survey_zou,PLS_yan,PLS_leihongjiang}. In this context, the
rise of ad-hoc or mesh-type communication technologies, such as device-to-device
(D2D) communications, presents new vulnerabilities. These technologies
can be leveraged by malicious users to jeopardize public safety, commit
crimes, coordinate terrorist activities, or illegally transmit confidential
trade information \cite{Surveillance_duanlingjie}. Addressing these
threats calls for the implementation of legitimate surveillance as
a critical component of wireless communication security. For example,
the National Security Agency of the United States launched the Terrorist
Surveillance Program in 2006 to proactively monitor and counter potential
threats \cite{1}. However, the rapidly increasing number of malicious
wireless devices over the past decade still poses growing concerns
over security threats. This highlights the need for a paradigm shift
from merely preventing conventional eavesdropping attacks to adopting
legitimate surveillance as a critical security measure \cite{Letter_duanlingjie}.

Physical-layer surveillance (i.e., monitoring) takes advantage of
the broadcast nature of wireless propagation \cite{PLS_yan}. As an
extension of secrecy rate, outage probability, and intercept probability,
etc., defined for physical layer security (PLS) performance analysis
\cite{MyTVT}, similar fundamental metrics have been adapted to evaluate
the performance of wireless surveillance strategies. Specifically,
the authors of \cite{Letter_duanlingjie} introduced the average eavesdropping
rate as a performance metric for monitoring, emphasizing that the
monitor operates effectively only when its achievable rate for intercepting
suspicious signals is greater than the suspicious communication rate.
Also, since the monitor overhears the suspicious signals for surveillance
purposes, the data rate of this received signal can be regarded as
the monitoring rate. Furthermore, similar to the secrecy rate, which
is the difference between the data rates of a legitimate user and
an eavesdropper, the relative monitoring rate (RMR) is defined as
the difference between the data rates of the legitimate surveillance
channel and the suspicious channel \cite{sun2022RISjam}. The probability
of a successful surveillance event, known as the surveillance success
probability (SSP) \cite{rotated_jamming_2019}, is defined as the
probability that the RMR is larger than a target threshold. For example,
the authors of \cite{TIFS_zhuhongbo} studied jamming power allocation
to maximize the RMR under an average transmit power constraint. To
solve their considered problems, both the bisection search and the
Lagrange duality method were applied.

A legitimate monitor can either silently receive suspicious signals
or engage in proactive eavesdropping through techniques such as spoofing
relaying \cite{Covert_2021} or cooperative jamming (CJ) \cite{zou2016},
where the CJ has been extensively investigated as a mean to degrade
the received signal quality of eavesdroppers \cite{jiangxiaointelligent,lijsrs}.
Specifically, the authors of \cite{jiangxiaointelligent} proposed
a threshold-based selection scheme to validate friendly jamming, and
formulated a subset of jammers with sufficiently strong channel quality
for selection. In fact, proactive monitoring is inspired by conventional
PLS methods to simultaneously counter eavesdropping and jamming attacks
\cite{spoofing_proactive_eavesdropping_2018}.When the channel gains
of the legitimate surveillance link are significantly weaker than
those of the suspicious communication link, passive monitoring becomes
inefficient because of its inability to decode suspicious messages.
In such situations, proactive monitoring via cooperative jamming emerges
as a more viable alternative. Specifically, the authors of \cite{TVTcorrespondence_huguojie}
studied two-phase relay-aided suspicious communication system and
proposed two strategies, namely ``passive eavesdropping first''
and ``jamming first'' to maximize the sum eavesdropping rate subject
to finite transmit power of the monitor.

While wireless surveillance has been regarded as a promising approach
to monitor suspicious communications, its effectiveness is still restricted
by uncontrollable radio environments in practice \cite{Surveillance_duanlingjie}.
To this end, reconfigurable intelligent surfaces (RISs) have emerged
as a powerful solution due to their unprecedented capability of manipulating
wireless propagation environments. Thus, extensive efforts have been
devoted to RIS-aided wireless surveillance. Specifically, the authors
of \cite{reactive_lizan} considered passive monitoring assisted by
a RIS, where signals transmitted from a suspicious transmitter to
a suspicious user were intercepted via a RIS-aided legitimate link.
In \cite{Zhaominjian2023_surveillance}, a full-duplex legitimate
monitor was studied in a proactive eavesdropping scenario utilizing
CJ, where the monitoring rate maximization problem was formulated
for three RIS deployment strategies. Then, a near-optimal performance
was achieved by jointly optimizing the receive and jamming beamforming
vectors at the legitimate monitor and the reflection coefficients
at the RIS. The authors of \cite{surveillance_bounded_error2022}
investigated a robust design for a RIS-aided wireless information
surveillance system with bounded channel errors. By jointly optimizing
the RIS phase shifts and receiver beamformer, the worst-case information
monitoring rate was maximized to improve surveillance performance.
In \cite{yizhi_clfirst}, a RIS-assisted cooperative jamming scheme
was proposed to combat suspicious communications. However, in this
scheme, the jammer was unable to obtain information from suspicious
communications.

Extensive research has been dedicated to performance analysis of monitoring
suspicious communications via CJ, as evidenced by the aforementioned
works. However, few studies have explored \emph{RIS-aided monitoring
with opportunistic selection among multiple jammers}. To address this
gap, in this paper, we study a RIS-aided wireless surveillance system
assisted by multiple jammers. The main contributions of this paper
are summarized as follows.
\begin{itemize}
\item First, we present two novel RIS-aided monitoring schemes with optimal
jammer selection based on different levels of channel state information
(CSI) available for RIS phase shift design. Unlike most existing studies
that assume perfect knowledge of all cascaded links at a central controller
(e.g., the monitor), we aim to achieve a tradeoff between optimal
phase shift design given full CSI and simpler randomly assigned phase
shifts by leveraging partial channel state information (CSI)%
. In the first scheme, referred to as RISLO, the phase shift design
relies on the CSI knowledge of the legitimate surveillance (LS) link.
In the second one, called RISCO, the CSI of the CJ link is employed
instead. These two schemes are compared with two corresponding benchmark
schemes, referred to as RISLR and RISCR that use random jammer selection
(RJS), both of which are also proposed for the first time in this
work.
\item We derive closed-form SSP expressions of the proposed schemes and
carry out an in-depth asymptotic analysis, leading to key insights.
Specifically, the RISLO and RISLR outperform RISCO and RISCR, respectively,
as they incorporate CSI from both LS and CJ links for phase shift
design and jammer selection. This indicates that monitoring performance
heavily depends on on the degree of CSI utilization, showing a fundamental
tradeoff between interaction/computation overhead and system performance.
Also, as jamming power increases, the monitoring performance reaches
a ceiling, where additional power consumption compensates for reduced
CSI requirements.
\item Moreover, we explore a special case where the jammers operate in an
almost passive manner and the number of RIS elements is asymptotically
large. The theoretical asymptotic analysis of this special case confirms
that RISLO outperforms RISCO, except when the monitoring channels
are significantly stronger than the suspicious channels.
\end{itemize}
The rest of the paper is organized as follows. Section II describes
the wireless surveillance system model. In Section III, we derive
closed-form SSP expressions of proposed schemes for different cases
of RIS phase shifts and jammer selection. Some asymptotic analysis
is further presented in Section IV. Numerical results are presented
in Section V. Finally, Section VI concludes the paper.

\emph{Notations}: Boldface lowercase letters and boldface uppercase
ones are used for vectors and matrices, respectively. For a complex
variable, $|\cdot|$ denotes its absolute value. For a complex vector,
$(\cdot)\trans$ and $(\cdot)\herm$ denote its respective transpose
and Hermitian transpose. Also, $\mathbb{C}^{N}$ and $\mathbb{C}^{M\times N}$
represent the complex-valued space of $N$-dimensional vectors and
the complex-valued space of $M$-by-$N$ matrices, respectively. Notations
$\sim$ and $\overset{\Delta}{=}$ stand for ``distributed as" and
``to be defined as", respectively. Besides, $n!$ represents the
factorial of a non-negative number $n$, $\diag(\boldsymbol{a})$
denotes a diagonal matrix with its diagonal elements given by $\boldsymbol{a}$,
$\text{arg}(\cdot)$ represents the phase of a complex number, i.e.,
$a=|a|\text{arg}(a)$, E$(\cdot)$ and Var$(\cdot)$ represent the
statistical expectation and variance operators, respectively, $G_{p,q}^{m,n}(\cdot)$
is the Meijer G-function {[}26, Eq. (9.301){]}, and $\Gamma(\cdot,\cdot)$
represents the upper incomplete gamma function, among which a special
case is the gamma function, noted as $\Gamma(0,\cdot)=\Gamma(\cdot)$,
where exists $\Gamma(n+1)=n!$ for a non-negative number $n$. Additionally,
$\binom{N}{n}$ is the number of possible cases to pick $n$ elements
from a set with $N$ elements.

\section{System Model}

\begin{figure}[tbh]
\centering {\includegraphics[scale=0.36]{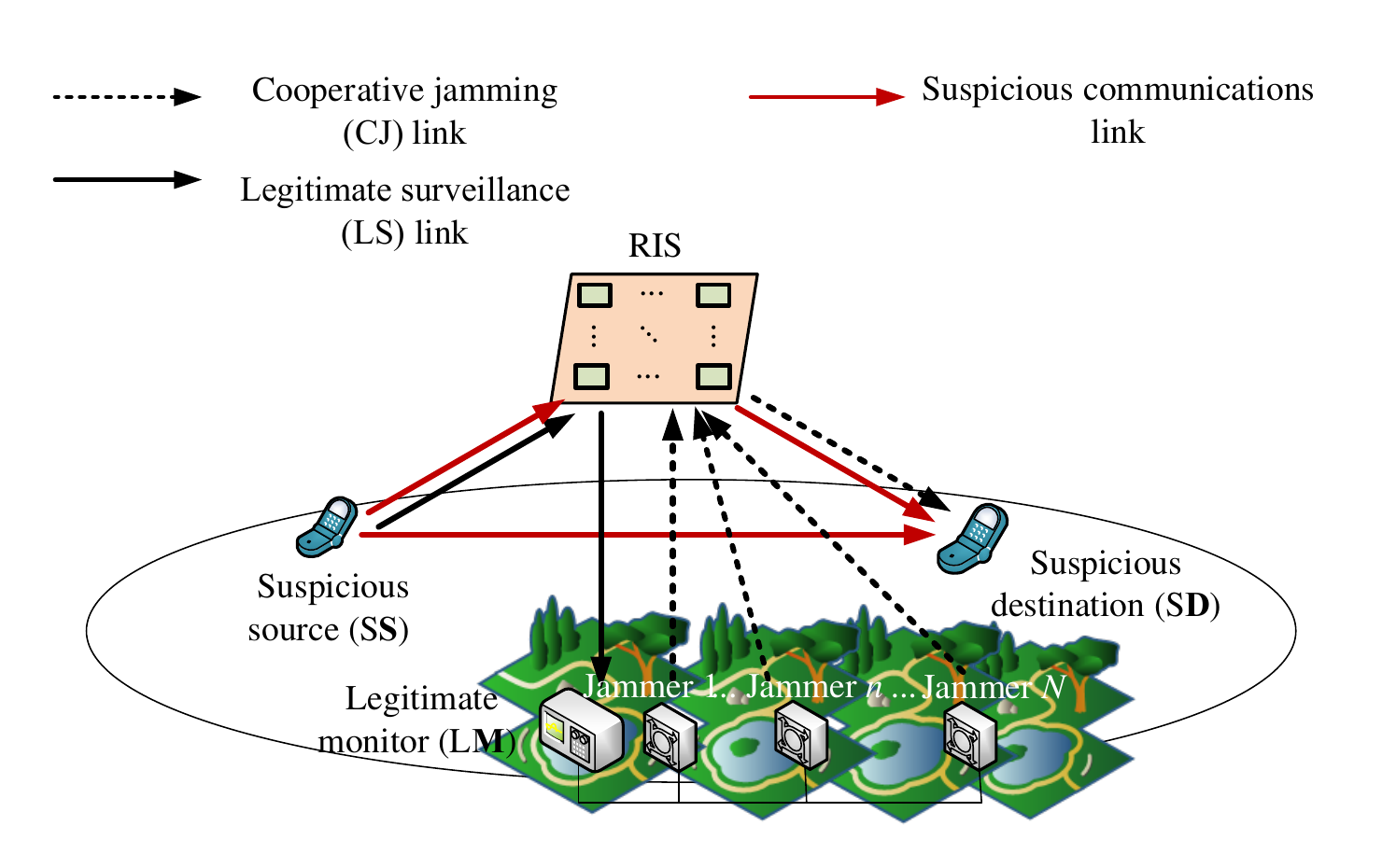}\\
 \caption{An RIS-aided wireless monitoring system assisted by multiple jammers.}
\label{Fig1}} 
\end{figure}

\subsection{RIS-Aided Monitoring System}

As illustrated in Fig. \ref{Fig1}, we consider a wireless monitoring
system consisting of a pair of suspicious source and destination (SS-SD),
a legitimate monitor (LM) including multiple distributed jammers,
and a RIS with $L$ co-located reflecting elements.\footnote{Each node is assumed to have a single antenna, while multi-antenna
nodes is left for future work.} The sets of RIS elements and jammers are denoted as $\mathcal{L}\overset{\Delta}{=}\{1,2,\cdots,L\}$
and $\mathcal{N}\overset{\Delta}{=}\{1,2,\cdots,N\}$, respectively.
One jammer is opportunistically selected to perform cooperative jamming
(CJ) to suspicious nodes based on a specific selection criterion.
When the SS transmits a signal to the SD at a power of $P_{\text{s}}$,
the LM can overhear the signal intended for the SD. Specifically,
the received signal at the LM is written as 
\begin{equation}
y_{\text{M}}=\sqrt{P_{\text{s}}}({\bf h}_{\text{RM}}\herm\boldsymbol{\boldsymbol{\Theta}}{\bf h}_{\text{SR}})x_{\text{s}}+n_{\text{M}},\label{equa(1)}
\end{equation}
where $x_{\text{s}}$ is the normalized symbol, i.e., $\text{E}\left(|x_{\text{s}}|^{2}\right)=1$,
${\bf h}_{\text{RM}}\herm\in\mathbb{C}^{1\times L}$ and ${\bf h}_{\text{SR}}\in\mathbb{C}^{L\times1}$
are channel coefficients of RIS-M and SS-RIS transmissions, respectively,
$\boldsymbol{{\boldsymbol{\Theta}}}$ is the reflection coefficient
diagonal matrix defined as $\boldsymbol{{\boldsymbol{\Theta}}}={\rm diag}([e^{-j\phi_{1}},\cdots,e^{-j\phi_{l}},\cdots,e^{-j\phi_{L}}])$,
where $\phi_{l}\in[0,2\pi)$ denotes the phase shift for each element
$l\in\mathcal{L}$, and $n_{\text{M}}$ is the additive white Gaussian
noise (AWGN) with zero mean and variance of $N_{0}$. In the considered
system model, it is assumed that the direct links between the LM,
the jammers and suspicious nodes are severely blocked. Consequently,
the LM and the jammers need to rely on the RIS to monitor the suspicious
nodes.

When the LM detects the presence of active suspicious nodes, a jammer
$n\in\mathcal{N}$ is selected to send a jamming signal $x_{\text{J}}$
deliberately at a power of $P_{\text{J}}$ (which is not the case
for passive monitoring) to decrease the signal-to-interference-plus-noise
ratio (SINR) at the SD. Thus, the received signal at the SD is given
by 
\begin{equation}
y_{\text{D,\emph{n}}}=\sqrt{P_{\text{s}}}({h}_{\text{SD}}+{\bf h}_{\text{RD}}\herm\boldsymbol{{\boldsymbol{\Theta}}}{\bf h}_{\text{SR}})x_{\text{s}}+\sqrt{P_{\text{J}}}({\bf h}_{\text{RD}}\herm\boldsymbol{{\boldsymbol{\Theta}}}{\bf h}_{n{\text{R}}})x_{\text{J}}+n_{\text{D}},\label{equa(2)}
\end{equation}
where ${h}_{\text{SD}}$, ${\bf h}_{\text{RD}}\herm\in\mathbb{C}^{1\times L}$,
${\bf {h}}_{\text{SR}}\in{{\mathbb{C}}^{L\times1}}$, and ${\bf {h}}_{n\text{R}}\in{{\mathbb{C}}^{L\times1}}$
are channel coefficients of SS-SD, RIS-SD, SS-RIS transmission, and
of the link from the $n$-th jammer to the RIS, respectively, $n\in\mathcal{N}$,
and $n_{\text{D}}$ is the AWGN with zero mean and variance of $N_{0}$
at the SD. We note that, with active jamming coming into play, the
signal model at the LM in \eqref{equa(1)} should include an interference
due to the RIS reflection of the jamming signal $x_{\text{J}}$. However,
since the jamming signal is already known by the LM, the reflected
interference can be effectively suppressed to a negligible level if
the LM can estimate the channel of the composite LM-RIS links \cite{cognitive_duanlingjie_2017}.
Thus, it is reasonable to adopt the simplified signal model in \eqref{equa(1)}
for further analysis. Also, we assume that the SS and SD are unaware
of the existence of the LM, and thus, do not employ any anti-eavesdropping
or anti-jamming methods \cite{two_suspicious_link}. As defined in\cite{MyTVT,rotated_jamming_2019},
the instantaneous capacity of SS-LM link is known as monitoring rate,
written as 
\begin{equation}
R_{\text{SM}}=\log_{2}(1+\gamma_{\text{s}}|{\bf h}_{\text{RM}}\herm\boldsymbol{{\boldsymbol{\Theta}}}{\bf h}_{\text{SR}}|^{2}),\label{equa(5)}
\end{equation}
while the instantaneous capacity of the SS-SD link is referred to
as suspicious rate, given by 
\begin{equation}
R_{\text{SD\emph{,n}}}=\log_{2}\left(1+\frac{\gamma_{\text{s}}|h_{\text{SD}}+{\bf h}_{\text{RD}}\herm\boldsymbol{{\boldsymbol{\Theta}}}{\bf h}_{\text{SR}}|^{2}}{\gamma_{\text{J}}|{\bf h}_{\text{RD}}\herm\boldsymbol{{\boldsymbol{\Theta}}}{\bf h}_{n\text{R}}|^{2}+1}\right),\label{equa(6)}
\end{equation}
where $\gamma_{\text{s}}=P_{\text{s}}/N_{0}$ and $\gamma_{\text{J}}=P_{\text{J}}/N_{0}$.

\subsection{Surveillance Success Probability}

In this section, we introduce the performance metric for physical-layer
surveillance. As discussed in Section \ref{sec:Introduction}, the
relative monitoring rate (RMR) is defined as the difference between
the monitoring rate and the suspicious rate, which is mathematically
expressed as \cite{Cell_free_monitor_successs_probability} 
\begin{equation}
R_{\text{M,\emph{n}}}=[R_{\text{SM}}-R_{\text{SD,\emph{n}}}]^{+}\mathpunct{,}\label{RMR}
\end{equation}
where $[x]^{+}={\rm max}\{0,x\}$. To define a successful monitoring
event, we consider that the RMR must be higher than a threshold $R_{\text{th}}$,
which represents the minimum target rate for the legitimate monitor
to decode successfully. The probability of this event is known as
surveillance success probability (SSP) and is given by 
\begin{equation}
P_{\text{ss}}=\Pr(R_{\text{M,\emph{n}}}>R_{\text{th}}),\label{equa(8)}
\end{equation}
where the criterion for choosing \emph{n} is specified in the following
section.

\section{Proposed RIS-Aided Monitoring Schemes and SSP Analysis}

While monitoring schemes using full CSI knowledge could theoretically
offer the best performance, they require an unlimited and unrealistic
amount of feedback, which is  practically infeasible. In this section,
we propose RIS-aided monitoring schemes where the RIS phase shifts
are optimized based on partial CSI. Our approach not only lowers the
complexity of phase shift optimization but also reduces the feedback
overhead associated with CSI acquisition. From \eqref{RMR}, it is
straightforward to see that, to maximize ${R_{\text{M}}}$, we can
increase ${R_{\text{SM}}}$ and/or decrease ${R_{\text{SD}}}$. However,
these two objectives are conflicting because ${R_{\text{SM}}}$ and
${R_{\text{SD}}}$ are interdependent due to their dependence on the
same phase shifts of the RIS, which affects the CSI of both links.
To address this, we consider two scenarios and propose proper strategies
accordingly. In the first case where the CSI of both LS link and CJ
link is available, we optimize the RIS phase shift to maximize ${R_{\text{SM}}}$
and perform jammer selection to minimize ${R_{\text{SD}}}$. In the
second case whereonly the CSI of the CJ link is known, we optimize
the RIS phase shifts and select a jammer to minimize ${R_{\text{SD}}}$.
Based on these strategies, we respectively propose two RIS-aided monitoring
schemes: RISLO and RISCO. In RISLO, the LM exploits the CSI of the
LS link combined with an optimally selected jammer. On the other hand,
in RISCO, the LM utilizes the CSI of the CJ link plus an optimally
selected jammer scheme. In addition, we provide a closed-form analysis
of the SSP for both schemes.

\subsection{RISLO: RIS phase optimization based on Legitimate surveillance channel with Optimal jammer selection}

\subsubsection{Phase Optimization and Jammer Selection}
In the RISLO scheme, the phase shifts are designed to maximize ${R_{\text{SM}}}$
given by \eqref{equa(5)}, i.e., to improve the average gain of SS-LM
transmission. Thus, the optimal phase shifts are given by 
\begin{equation}
\phi_{l}^{{\text{RISLO}}}={\rm arg(}h_{\text{R}_{l}\text{M}}^{*}{\rm )}{\rm +arg(}h_{\text{SR}_{l}}),\quad\forall l\in\mathcal{L}.\label{equa(3)}
\end{equation}
Once the phase shifts are determined, we opportunistically choose
the jammer that minimizes $R_{\textrm{SD}}$. To this end, we rewrite
\eqref{equa(6)} as
\begin{equation}
{R_{\text{SD},n}}={\log_{2}}\left(1+\frac{\gamma_{\text{s}}Y^{\text{RISLO}}}{\gamma_{\text{J}}Q_{n}^{\text{RISLO}}+1}\right)\label{equa(22)}
\end{equation}
where $Y^{\text{RISLO}}=|{h}_{\text{SD}}+{\bf h}_{\text{RD}}\herm{\boldsymbol{\Theta}^{\text{RISLO}}}{\bf {h}}_{\text{SR}}|^{2}$
is the cascaded channel gain of the suspicious link, and $Q_{n}^{\text{RISLO}}=|{\bf h}_{\text{RD}}\herm\boldsymbol{\Theta}^{\text{RISLO}}{\bf {h}}_{n\mathit{\textrm{R}}}|^{2}$
denotes the gain of RIS-aided CJ channels, wherein $\boldsymbol{\Theta}^{\text{RISLO}}=\diag(e^{-j{\rm [arg(}{\bf h}_{\text{\ensuremath{\text{R}_{l}\text{M}}}}\herm{\rm )}{\rm +arg(}\boldsymbol{h}_{\text{SR}_{l}}{\rm )]}})$%
. It is obvious that to reduce $R_{\text{SD},n}$, we select the optimal
jammer as%
\begin{equation}
\begin{aligned}m={\rm {arg}}\mathop{{\rm max}}\limits_{n\in\mathcal{N}}Q_{n}^{\text{RISLO}}.\end{aligned}
\label{equa(21)}
\end{equation}

\subsubsection{SSP Analysis of RISLO}

After optimizing the phase shifts and selecting a jammer, we now proceed
to analyze the SSP performance of the RISLO scheme. To start with,
we derive the necessary statistical distributions to facilitate subsequent
derivations. First, we remark that, in the RISLO scheme, the phase
shifts $\boldsymbol{\Theta}^{\text{RISLO}}$ in $Q_{n}^{\text{RISLO}}$
exhibit equivalent properties to random phase shifts due to the independence
of the LS link and CJ links (see the Appendix for further discussions)
.It is also clear from the Appendix that the suspicious channel gain
belongs to the same case. Then, both $Y^{\text{RISLO}}$ and $Q_{n}^{\text{RISLO}}$
follow exponential distributions.  $Q_{n}^{\text{RISLO}}$ follows
an exponential distribution given by \eqref{equa(14)-1} in the Appendix.
Similarly, given that $Y^{\text{RISLO}}$ is independently but not
necessarily identically distributed, its cumulative density function
(CDF) is written as 
\begin{equation}
F_{Y^{\text{RISLO}}}(q)=1-{e^{-\frac{q}{\sigma_{\text{SD}}^{2}+L\sigma_{\text{RD}}^{2}\sigma_{\text{SR}}^{2}}}}.\label{equa(14)}
\end{equation}
By defining $W=|{\bf h}_{\text{RM}}\herm{\boldsymbol{\Theta}^{\text{RISLO}}}{\bf {h}}_{\text{SR}}|$
and applying \eqref{equa(3)}, we can simplify $W$ to 
\begin{equation}
W=\sum\limits_{l=1}^{L}{|h_{\text{R}_{l}\text{M}}||h_{\text{SR}_{l}}|},\label{defineW1}
\end{equation}
where $h_{\text{R}_{l}\text{M}}$ and $h_{\text{SR}_{l}}$ are modeled
as independent zero-mean complex Gaussian random variables with variances
of $\sigma_{\text{RM}}^{2}$ and $\sigma_{\text{SR}}^{2}$, respectively.
These assumptions are based on independently and identically distributed
Rayleigh fading channels from different reflecting elements of the
RIS. Using the Laguerre series approximation and following the existing
literature on RISs \cite{IRS_NOMA}\cite{performance_active}, we
approximate the CDF of $W$ as a Gamma distribution given by 
\begin{equation}
\Pr(W\le w)=1-\frac{\Gamma(\lambda,\frac{w}{w_{1}})}{\Gamma(\lambda)},\label{equa(10)}
\end{equation}
where the shape and scale parameters are given by
\begin{equation}
\lambda=\frac{\text{E}^{2}(W)}{\text{Var}(W)}=\frac{\pi^{2}L}{16-\pi^{2}},\quad w_{1}=\frac{\text{Var}(W)}{\text{E}(W)}.\label{equa(11)}
\end{equation}
Here $\text{E}(W)$ and ${\text{Var}}(W)$ denote the mean and variance
of $W$, respectively. In the above, we have used the moment-match
method, which effectively models positive random variables whose PDF
has a single maximum and fast decaying tails \cite{Stochastic}. The
statistical parameters are derived as 
\begin{equation}
\text{E}(W)=\frac{\pi L}{16}\sigma_{\text{RM}}\sigma_{\text{SR}},\label{equa(12)}
\end{equation}
and 
\begin{equation}
\begin{aligned}{\text{Var}}(W) & =\pi L\left[\text{E}(|h_{\text{RM}}|^{2}|h_{\text{SR}}|^{2})-\text{E}^{2}(|h_{\text{RM}}||h_{\text{SR}}|)\right]\\
 & =\pi L\sigma_{\text{RM}}^{2}\sigma_{\text{SR}}^{2}(1-\frac{\pi^{2}}{16}),
\end{aligned}
\label{equa(13)}
\end{equation}
which completes the statistical characterization of the channel gain
from LS link.

Combining \eqref{equa(22)} and \eqref{equa(21)}, we know that ${R_{\text{SD}}^{\text{RISLO}}=R_{\text{SD},m}}$.
Letting $V=\frac{2^{R_{\text{th}}}Y_{}^{\text{RISLO}}}{\gamma_{\text{J}}\mathop{{\rm max}}\limits_{n\in\mathcal{N}}Q_{n}^{\text{RISLO}}+1}$,
we obtain the CDF of $V$ as 
\begin{equation}
\begin{aligned} & {F_{V}}(v)=\Pr\left(V\leq v\right)\\
= & \int_{0}^{\infty}{\frac{1}{\Xi}{e^{-\frac{y}{\Xi}}}\left[1-\prod\limits_{n\in\mathcal{N}}{\left({1-{e^{-\frac{\frac{2^{R_{\text{th}}}}{\gamma_{\text{J}}}y-1}{v(L\sigma_{n_{\text{R}}}^{2}\sigma_{_{\text{RD}}}^{2})}}}}\right)}\right]}{\rm d}y\\
= & \int_{0}^{\infty}{\frac{e^{\frac{1}{v(L\sigma_{n_{\text{R}}}^{2}\sigma_{_{\text{RD}}}^{2})}}}{\Xi}{e^{-\frac{y}{\Xi}}}\sum\limits_{t=1}^{2^{N}-1}{{(-1)}^{|{J_{t}}|+1}e^{-\sum\limits_{{J_{t}}}{\frac{y}{v(L\sigma_{n_{\text{R}}}^{2}\sigma_{_{\text{RD}}}^{2})}}}}}{\rm d}y\\
= & \sum\limits_{n=1}^{N}{\binom{N}{n}\frac{(-1)^{n+1}v{e^{\frac{1}{v(L\sigma_{n_{\text{R}}}^{2}\sigma_{_{\text{RD}}}^{2})}}}}{v+n\delta_{1}}},
\end{aligned}
\label{equa(23)}
\end{equation}
the CSI for $N$ different jammers is considered independent. ${J_{t}}$
represents the $t$-th non-empty subcollection of the jammer set $\mathcal{N}$,
and $\delta_{1}=\frac{2^{R_{\text{th}}}\Xi}{\gamma_{\text{J}}(L\sigma_{n_{\text{R}}}^{2}\sigma_{_{\text{RD}}}^{2})}$.
Besides, $|{J_{t}}|$ denotes the cardinality of the set $J_{t}$,
and $\binom{N}{n}$ is the number of all possible subcollections satisfying
$|{J_{t}}|=n$. By substituting \eqref{equa(10)} and \eqref{equa(23)}
into \eqref{equa(22)}, the SSP of the RISLO scheme can be derived
as 
\begin{equation}
\begin{aligned}{P_{\text{ss}}^{\text{RISLO}}}= & \Pr({R_{\text{SM}}}-{R_{\text{SD}}^{\text{RISLO}}}>{R_{\text{th}}})\\
= & \int_{0}^{\infty}{\frac{{f_{W_{1}}\left(\sqrt{v+\beta}\right)}}{2\sqrt{v+\beta}}{F_{V}}\left(v\right){e^{\frac{1}{v(L\sigma_{n_{\text{R}}}^{2}\sigma_{_{\text{RD}}}^{2})}}}}{\rm d}v.
\end{aligned}
\label{equa(24)}
\end{equation}
Substituting \eqref{equa(23)} into \eqref{equa(24)}, and capitalizing
on the Gaussian-Chebyshev quadrature \cite{book3}, the SSP of the
RISLO scheme is given by \emph{\eqref{equa(25)}} shown at the top
of the page, where $\theta_{k}=\cos\left(\frac{2k-1}{2K}\pi\right)$,
$\tau_{k}=\frac{(\theta_{k}+1)\pi}{4}$, and $K$ is accuracy versus
complexity parameter. 
\begin{figure*}[t]
\begin{equation}
{\footnotesize \begin{aligned}{P_{\text{ss}}^{\text{RISLO}}}=\frac{\pi^{2}}{4K}\sum_{k=1}^{K}{\sum\limits_{n=1}^{N}{\binom{N}{n}\frac{(-1)^{n+1}\sqrt{1-\theta_{k}^{2}}\sec^{2}{\tau_{k}}\tan{\tau_{k}}{\left(\sqrt{\tan{\tau_{k}}+\beta}\right)}^{\lambda-2}{e^{-\frac{\sqrt{\tan{\tau_{k}}+\beta}}{w_{1}}}}{e^{\frac{1}{\tan{\tau_{k}}(L\sigma_{n_{\text{R}}}^{2}\sigma_{_{\text{RD}}}^{2})}}}}{2{{w_{1}}^{\lambda}}\left({\lambda-1}\right)!(\tan{\tau_{k}}+n\delta_{1})}}}\end{aligned}
}\label{equa(25)}
\end{equation}
\hrule
\end{figure*}

\begin{figure*}[t]
\begin{equation}
{P_{\text{ss}}^{\text{RISLR}}}=\frac{\pi^{2}}{4K}\sum_{k=1}^{K}{\frac{\sqrt{1-\theta_{k}^{2}}\sec^{2}{\tau_{k}}\tan{\tau_{k}}{\left(\sqrt{\tan{\tau_{k}}+\beta}\right)}^{\lambda-2}{e^{-\frac{\sqrt{\tan{\tau_{k}}+\beta}}{w_{1}}}}e^{\frac{1}{\tan{\tau_{k}}(L\sigma_{n_{\text{R}}}^{2}\sigma_{_{\text{RD}}}^{2})}}}{2{{w_{1}}^{\lambda}}\left({\lambda-1}\right)!(\tan{\tau_{k}}+\delta_{1})}}\label{RISLR}
\end{equation}
\hrule
\end{figure*}

To highlight the performance gain from jammer selection, we adopt
the RISLR as a benchmark scheme corresponding to RISLO. The RISLR
adopts an equal-probability selection from the jammer set $\mathcal{N}$
instead of \eqref{equa(21)}. The SSP expression of RISLR is written
as \eqref{RISLR} shown at the top of the page for comparison. We
skip the derivation for brevity as it remains the same as in this
section.

\begin{rem}[Asymptotic analysis with high jamming SNR]
\emph{When the jamming power $P_{\text{J}}$ increases indefinitely,
the parameter $\delta_{1}$ approaches zero. By comparing \eqref{equa(25)}
and \eqref{RISLR}, and considering the fact that $\sum\limits_{n=0}^{N}{\binom{N}{n}(-1)^{n}}=0$,
it follows that the two expressions converge to the same value as
$\delta_{1}\rightarrow0$. This indicates that the benefit of CSI-based
jammer selection becomes marginal in high-power jamming scenarios.
The reasons is that all CJ links are in good conditions, diminishing
the impact of jammer selection. In contrast, when the jamming power
decreases, $\delta_{1}$ becomes larger, leading to a significant
performance gap between RISLO and RISLR. In this regime, the RISLO
scheme outperforms RISLR by leveraging CSI to select the most effective
jammer.}
\end{rem}

\subsection{RISCO: RIS phase optimization based on Cooperative jamming channel
with Optimal jammer selection}

\subsubsection{Phase Optimization and Jammer Selection}

In the RISCO scheme, the RIS phase shifts are designed to maximize
the denominator in \eqref{equa(6)} for a given jammer, relying on
the CSI of CJ links. For an arbitrary jammer $n$, the desired phase
shifts aim to maximize the average channel gain between the monitor
and the suspicious receiver, and thus are given by 
\begin{equation}
\phi_{l\emph{,n}}^{\text{ RISCO}}={\rm arg(}h_{\text{R}_{l}\text{D}}^{*}{\rm )}{\rm +arg(}h_{n\text{R}_{l}}{\rm )},\quad\forall l\in\mathcal{L}\label{equa(4)}
\end{equation}
which gives
\begin{equation}
{Q_{n}^{\text{RISCO}}}=|{\bf h}_{\text{RD}}\herm{\boldsymbol{\Theta}}{\bf {h}}_{n\text{R}}|^{2}=\sum\limits_{l=1}^{L}{|h_{\text{R}_{l}\text{D}}||h_{\text{\emph{n}R}_{l}}|}.\label{defineW1-1}
\end{equation}
In the RISCO scheme, the optimal jammer is selected to minimize $R_{\text{SD},n}$
by maximizing the channel gain of the RIS-aided CJ links. Specifically,
the optimal jammer is determined as 
\begin{equation}
\begin{aligned}c={\rm {arg}}\mathop{{\rm max}}\limits_{n\in\mathcal{N}}{Q_{n}^{\text{RISCO}}}.\end{aligned}
\label{equa(34)}
\end{equation}

\subsubsection{SSP Analysis of RISCO}

Combining \eqref{equa(4)} with \eqref{equa(34)} and by defining
$\boldsymbol{\Theta}^{\text{RISCO}}={\textstyle \diag}(e^{-j{\rm [arg(}\boldsymbol{h}_{\text{R}_{l}\text{D}}^{H}{\rm )}{\rm +arg(}\boldsymbol{h}_{c\text{R}_{l}}{\rm )]}})$%
, ${Q_{n}^{\text{RISCO}}}$ given by \eqref{defineW1-1} is characterized
analogously as $W$, and the same steps are followed to derive the
distribution of ${Q_{n}^{\text{RISCO}}}$. From \eqref{equa(10)},
the CDF of ${Q_{n}^{\text{RISCO}}}$ is expressed as 
\begin{equation}
\Pr\bigl({Q_{n}^{\text{RISCO}}}\le w\bigr)=1-\frac{\Gamma(\lambda_{2},\frac{w}{w_{2,n}})}{\Gamma(\lambda_{2})},\label{equa(26)}
\end{equation}
where 
\begin{equation}
\lambda_{2}=\frac{\pi^{2}L}{16-\pi^{2}}=\lambda,\label{equa(27)}
\end{equation}
wherein $\lambda$ is also given in \eqref{equa(11)} and 
\begin{equation}
w_{2,n}=\frac{(16-\pi^{2})\sigma_{\text{RD}}\sigma_{n\text{R}}}{4\pi}.\label{equa(28)}
\end{equation}
Similar to the RISLO scheme, the phase shift of RIS in the RISCO scheme
given by \eqref{equa(4)} is random for the LS link. Then, $Y^{\text{RISCO}}=|{h}_{\text{SD}}+{\bf h}_{\text{RD}}\herm{\boldsymbol{\Theta}^{\text{RISCO}}}{\bf {h}}_{\text{SR}}|^{2}$
is the same as that of the RISLO scheme. Letting $Z_{1}=|{\bf h}_{\text{RM}}\herm{\boldsymbol{\Theta}^{\text{RISCO}}}{\bf {h}}_{\text{SR}}|^{2}$,
$Z_{1}$ follows an exponential distribution with its CDF given by
\begin{equation}
F_{Z_{1}}(z)=1-{e^{-\frac{z}{L\sigma_{_{\text{SR}}}^{2}\sigma_{_{\text{RM}}}^{2}}}}.\label{equa(29)}
\end{equation}
Letting $G=\frac{Y^{\text{RISCO}}}{Z_{1}-\beta}$, where $\beta=\frac{2^{R_{\text{th}}}-1}{\gamma_{\text{s}}}$,
we derive the CDF of $G$ as 
\begin{equation}
{F_{G}}(g)=\int_{0}^{\infty}{\frac{1}{\Xi}{e^{-\frac{y}{\Xi}}}{e^{-\frac{\frac{y}{g}+\beta}{L\sigma_{{\text{SR}}}^{2}\sigma_{_{\text{RM}}}^{2}}}}}{\rm d}y=\frac{ge^{-\frac{\beta}{L\sigma_{{\text{SR}}}^{2}\sigma_{_{\text{RM}}}^{2}}}}{g+\delta_{2}},\label{equa(32)}
\end{equation}
where $\delta_{2}=\frac{\Xi}{L\sigma_{{\text{SR}}}^{2}\sigma_{_{\text{RM}}}^{2}}$.
Similar to \eqref{equa(22)}, we obtain the suspicious rate of the
RISCO scheme as 
\begin{equation}
{R_{\text{SD}}^{\text{RISCO}}}=R_{\text{SD},c}={\log_{2}}\left(1+\frac{\gamma_{\text{s}}Y^{\text{RISCO}}}{\gamma_{\text{J}}\mathop{{\rm max}}\limits_{n\in\mathcal{N}}Q_{n}^{\text{RISCO}}+1}\right).\label{equa(35)}
\end{equation}
By letting $T=\mathop{{\rm max}}\limits_{n\in\mathcal{N}}Q_{n}^{\text{RISCO}}$,
then the PDF of $T$ can be obtained as \eqref{equa(36)} shown at
the top of the page, where the generalized multinomial theorem is
utilized, and ${P_{q,n}}$ represents the $q$-th non-empty subcollection
of the jammer set $\{\mathcal{N}-n\}$, $|{P_{q,n}}|$ denote the
cardinality of the set $P_{q,n}$. Besides, note that the set $\mathcal{S}=\{\left(n_{1},n_{2},\ldots,n_{\lambda}\right)|\sum_{p=1}^{\lambda}n_{p}=|{P_{q,n}}|\}$,
$A_{1}=\frac{\prod_{k=1}^{\lambda}{\frac{1}{((k-1)!)^{n_{k}}}}}{\prod_{p=1}^{\lambda}n_{p}!}$,
$B_{1}=\sum_{p=1}^{\lambda}{n_{p}(p-1)}$.
\begin{figure*}
\begin{eqnarray}
{f_{T}}(t) & = & \sum\limits_{n=1}^{N}{\frac{t^{\lambda-1}}{{(\mu_{n})}^{\lambda}\left({\lambda-1}\right)!}{e^{-\frac{t}{\mu_{n}}}}\prod\limits_{m\in\{\mathcal{N}-n\}}\left(1-{e^{-\frac{t}{\mu_{m}}}}\left({\sum\limits_{k=1}^{\lambda-1}{\frac{t^{k}}{\mu_{m}^{k}}}}\right)\right)}\nonumber \\
 & = & \sum\limits_{n=1}^{N}{\frac{1}{\left({\lambda-1}\right)!}}\left(1+\sum\limits_{q=1}^{{2^{N-1}}-1}{{(-1)}^{|{P_{q,n}}|}}\sum\limits_{m\in P_{q,n}}{e^{-\frac{t}{\mu_{n}}-{\frac{|{P_{q,n}}|t}{\mu_{m}}}}}\sum_{\mathcal{S}}{\frac{A_{1}}{{(\mu_{m})}^{B_{1}}}t^{B_{1}+\lambda-1}}\right).\label{equa(36)}
\end{eqnarray}

\hrule
\end{figure*}

Then, the SSP of the RISCO scheme can be derived as 
\begin{equation}
\begin{aligned}{P_{\text{ss}}^{\text{RISCO}}} & =\Pr({R_{\text{SM}}}-{R_{\text{SD}}^{\text{RISCO}}}>{R_{\text{th}}})\\
 & =\int_{\frac{1}{\sqrt{\gamma_{\text{J}}}}}^{\infty}{\frac{\sqrt{\gamma_{\text{J}}}{f_{T}\left(\sqrt{\frac{\gamma_{\text{J}}g^{2}-1}{2^{R_{\text{th}}}}}\right)}}{\sqrt{2^{R_{\text{th}}}}}{F_{G}}\left(g^{2}\right)}{\rm d}g.
\end{aligned}
\label{equa(37)}
\end{equation}
Substituting \eqref{equa(32)} and \eqref{equa(36)} into \eqref{equa(37)},
the closed-form SSP expression of the RISCO scheme given by \eqref{equa(38)}
shown at the top of next page, where $\beta=\frac{2^{R_{\text{th}}}-1}{\gamma_{\text{s}}}$,
$G_{p,q}^{m,n}(\cdot)$ is the Meijer G-function \cite[Eq. (9.301)]{book2},
and the result of \cite[Eq. (3.389-2)]{book2} is used. 
\begin{figure*}
\begin{equation}
{P_{\text{ss}}^{\text{RISCO}}}=\frac{e^{-\frac{\beta}{L\sigma_{{\text{SR}}}^{2}\sigma_{_{\text{RM}}}^{2}}}}{2\sqrt{\pi}\left({\lambda-1}\right)!}\sum\limits_{n=1}^{N}{\frac{1}{{(\mu_{n})}^{\lambda}}}\left(1+\sum\limits_{q=1}^{{2^{N-1}}-1}\sum\limits_{m\in P_{q,n}}{\frac{{(-1)}^{|{P_{q,n}}|}}{|{P_{q,n}}|!}}\sum_{\mathcal{S}}{A_{1}\delta_{2}^{\frac{\lambda+B_{1}}{2}}G_{1,3}^{3,1}\left(\frac{\left(\mu_{n}+|{P_{q,n}}|\mu_{m}\right)^{2}\delta_{2}}{4}\Bigg|\begin{array}{c}
-\frac{\lambda+B_{1}}{2}\\
-\frac{\lambda+B_{1}}{2},0,\frac{1}{2}
\end{array}\right)}\right).\label{equa(38)}
\end{equation}

\hrule
\end{figure*}

\section{Asymptotic Analysis in Useful Special Cases}

Since the intricate SSP expressions involving the special functions
(e.g., Meijer G-function) do not facilitate direct performance comparisons
between different schemes, we present an asymptotic analysis that
accounts for practical limitations. This approach offers useful insights,
which help system designers select the most suitable scheme for a
given situation. Several proposed schemes have shown obviously competitive
performance exploiting proactive monitors, but it is not fair to compare
with passive monitoring because of the extra power consumption and
implementation complexity of full-duplex devices. Besides, it is preferred
to consider the monitor to be invisible to the suspicious pairs, where
jamming power control should be focused as well. Otherwise, the suspicious
receiver becomes aware and takes anti-jamming measures, causing all
our jammers to suffer performance loss.

Let us focus on the special scenario where the jammers operate with
minimal transmit power, i.e., $P_{\text{J}}\rightarrow0$. That is,
we aim to characterize the asymptotic behavior of the SSP under nearly
passive jamming conditions In this case, the derived expressions are
simplified, allowing for clearer insights into system performance.
As $P_{\text{J}}\rightarrow0$, \eqref{equa(22)} reduces to 
\begin{equation}
{R_{\text{SD}}^{\text{RISLO},{\text{Pas}}}}={\log_{2}}\left(1+\gamma_{\text{s}}Y_{2}\right),\label{equa(16)}
\end{equation}
where the superscript ``Pas" denotes the passive monitoring approximation,
implying low jamming SNR. For further analytical insights, we set
${R_{\text{th}}}=0$, which does not change the general trend of SSP
functions. By combining \eqref{equa(25)} and \eqref{equa(16)}, the
asymptotic SSP of the RISLO scheme can be expressed as 
\begin{align}
P_{\text{ss}}^{\text{RISLO},{\text{Pas}}} & =\Pr\left(W_{1}^{2}>Y_{2}\right)\nonumber \\
 & =\left(\frac{4{w_{1}}^{2}}{L\sigma_{n_{\text{R}}}^{2}\sigma_{_{\text{RD}}}^{2}}\right)^{\frac{1}{4}-\frac{\lambda}{2}}e^{\frac{L\sigma_{n_{\text{R}}}^{2}\sigma_{_{\text{RD}}}^{2}}{8{w_{1}}^{2}}}\mathcal{W}_{{\frac{1}{4}-\frac{\lambda}{2}},-\frac{1}{4}}\left({\frac{L\sigma_{n_{\text{R}}}^{2}\sigma_{_{\text{RD}}}^{2}}{4{w_{1}}^{2}}}\right)\label{RISLO}
\end{align}
where $\mathcal{W}_{a,b}$(.) is the Whittaker function, as defined
in \cite[Eq. (9.222-2)]{book2}. To derive the above we have used
the result of \cite[Eq. (3.462-1)]{book2} and the definition of parabolic
cylinder functions \cite[Eq. (9.240)]{book2}. However, due to the
complexity of expressions containing the Whittaker function, it is
still challenging to gain more insights. To highlight the performance
gains enabled by the RIS, we examine the asymptotic behavior for a
large number of RIS elements, i.e., $L\rightarrow\infty$. As $\frac{L\sigma_{n_{\text{R}}}^{2}\sigma_{_{\text{RD}}}^{2}}{4{w_{1}}^{2}}$
in \ref{RISLO} approaches infinity, the Whittaker function can be
approximated using Watson’s lemma \cite{book2}, which results in
\begin{equation}
\mathrm{W}_{a,b}(z)\sim\mathrm{e}^{-\frac{z}{2}}z^{a}\sum_{m=0}^{\infty}(-1)^{m}\frac{\left(\frac{1}{2}-a+b\right)_{m}\left(\frac{1}{2}-a-b\right)_{m}}{n!z^{m}},
\end{equation}
where ${(a_{p})}_{k}=\frac{\Gamma(a_{p}+k)}{\Gamma(a_{p})}$, $a={\frac{1}{4}-\frac{\lambda}{2}}$
and $b=-\frac{1}{4}$, also $z=\frac{L\sigma_{n_{\text{R}}}^{2}\sigma_{_{\text{RD}}}^{2}}{4{w_{1}}^{2}}$.
Then, \eqref{RISLO} simplifies to 
\begin{equation}
\begin{aligned}{P_{\text{ss}}^{\text{RISLO},{\text{Pas}}}} & =\sum_{m=0}^{\infty}(-1)^{m}\frac{\left(\frac{1}{2}-a+b\right)_{m}\left(\frac{1}{2}-a-b\right)_{m}}{n!z^{m}}\\
 & \overset{(a)}{=}\mathcal{O}(e^{-\frac{1}{z}}),
\end{aligned}
\end{equation}
where $(a)$ follows from the Taylor expansion, and $\mathcal{O}(e^{-\frac{1}{z}})$
means a similar asymptotic behavior as $e^{-\frac{1}{z}}$ given $L\rightarrow\infty$,
i.e., $z\rightarrow\infty$.

By combining \eqref{equa(25)} and \eqref{equa(16)}, along with \eqref{equa(14)}
and \eqref{equa(29)}, the asymptotic SSP of the RISCO scheme is expressed
as 
\begin{equation}
\begin{array}{l}
\begin{aligned}P_{\text{ss}}^{\text{RISCO},{\text{Pas}}} & =\Pr\left(Z_{1}>Y_{2}\right)\\
 & =\frac{1}{1+\delta_{2}},
\end{aligned}
\end{array}\label{RISCO}
\end{equation}
where $\delta_{2}=\frac{\sigma_{\text{SD}}^{2}+L\sigma_{\text{RD}}^{2}\sigma_{\text{SR}}^{2}}{L\sigma_{{\text{SR}}}^{2}\sigma_{_{\text{RM}}}^{2}}$.

\noindent{}
\begin{rem}[Performance gain from RIS in RISLO and RISCO]
\emph{The performance improvements offered by RIS can be analyzed
by comparing \eqref{RISLO} and \eqref{RISCO}. Notably, with an asymptotically
large number of RIS elements, the SSP performance of the }RISLO\emph{
and }RISCO\emph{ schemes is significantly different. When $L$ becomes
sufficiently large, $P_{\text{ss}}^{\text{RISLO}}$ approaches one
at an exponential rate. In contrast, although $P_{\text{ss}}^{\text{RISCO},{\text{Pas}}}$
is also an increasing function of $L$, it converges only to a constant
between zero and one. To facilitate the analysis, we define the monitoring
to suspicious ratio (MSR) as the ratio of average gains between the
monitoring channel and suspicious channel, i.e., $\zeta_{\text{MSR}}=\frac{\sigma_{n_{1}}^{2}}{\sigma_{n_{2}}^{2}}$
where $n_{1}\in\{{\text{\text{\ensuremath{{\textstyle RM}}}}},{n\text{\text{\ensuremath{{\textstyle R}}}}}\}$
and $n_{2}\in\{{\text{\text{\ensuremath{{\textstyle SR}}}}},{\text{\text{\ensuremath{{\textstyle RD}}}}}\}$.
This ratio provides a comparative measure of the channel quality between
the suspicious and monitoring links. As $L\rightarrow\infty$, we
obtain $\delta_{2}\rightarrow\frac{1}{\zeta_{\text{MSR}}}$ which
leads to the asymptotic bound: $P_{\text{ss}}^{\text{RM-CSICJ},{\text{Pas}}}\rightarrow\frac{\zeta_{\text{MSR}}}{\zeta_{\text{MSR}}+1}$.
This result indicates that the RISCO scheme achieves a strong theoretical
performance bound only when the LS and CJ links significantly outperform
the suspicious links. However, we note that this condition is not
always guaranteed in practical implementations.}
\end{rem}

\section{Simulation Results and Discussions}

\begin{figure}[tbh]
\centering {\includegraphics[scale=0.58]{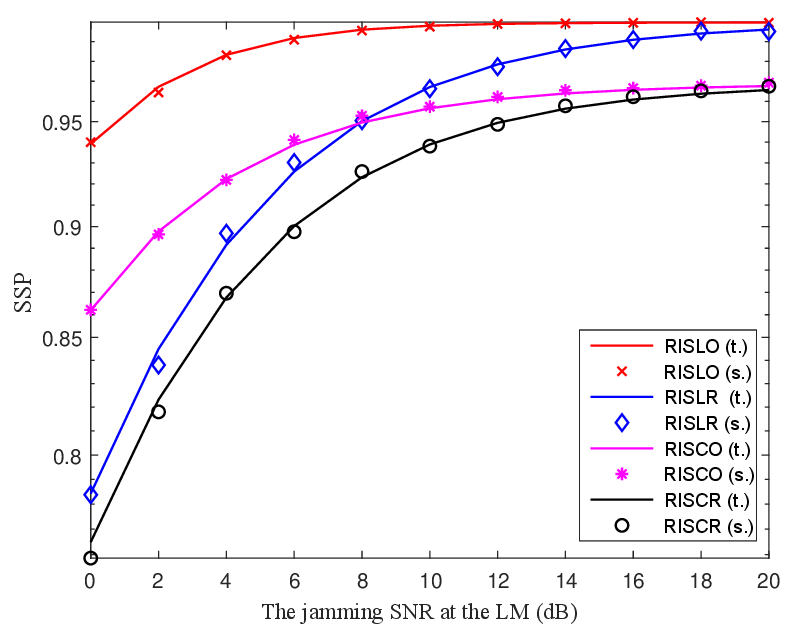}\\
 \caption{Surveillance outage probability of RISLR, RISLO, RISCR, and RISCO
schemes versus jamming SNR, where \textquotedblleft t." and \textquotedblleft s."
stand for theoretical and simulation results, respectively.}
\centering\label{Fig2}} 
\end{figure}

\begin{figure}[tbh]
\centering {\includegraphics[scale=0.58]{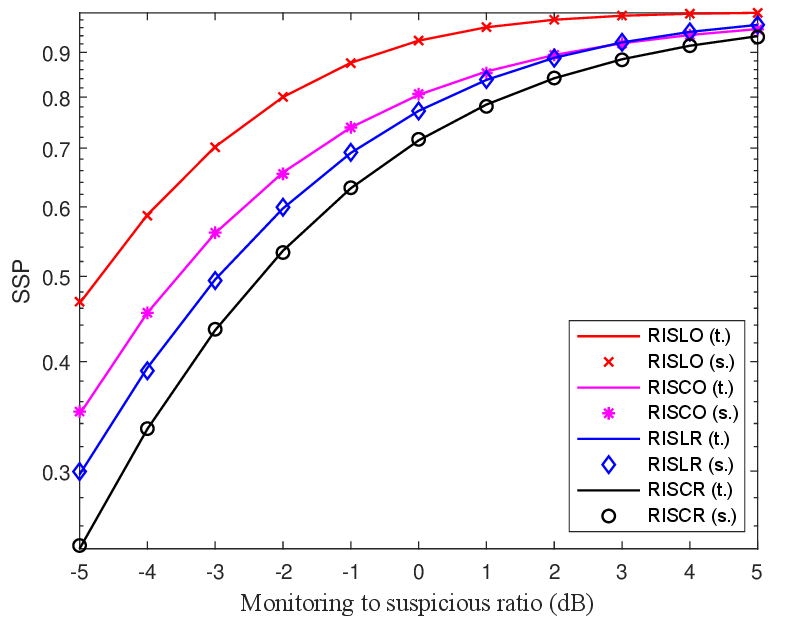}\\
 \caption{Surveillance success probability of RISLR, RISLO, RISCR, and RISCO
schemes versus MSR.}
\centering\label{Fig3}} 
\end{figure}

\begin{figure}[tbh]
\centering {\includegraphics[scale=0.58]{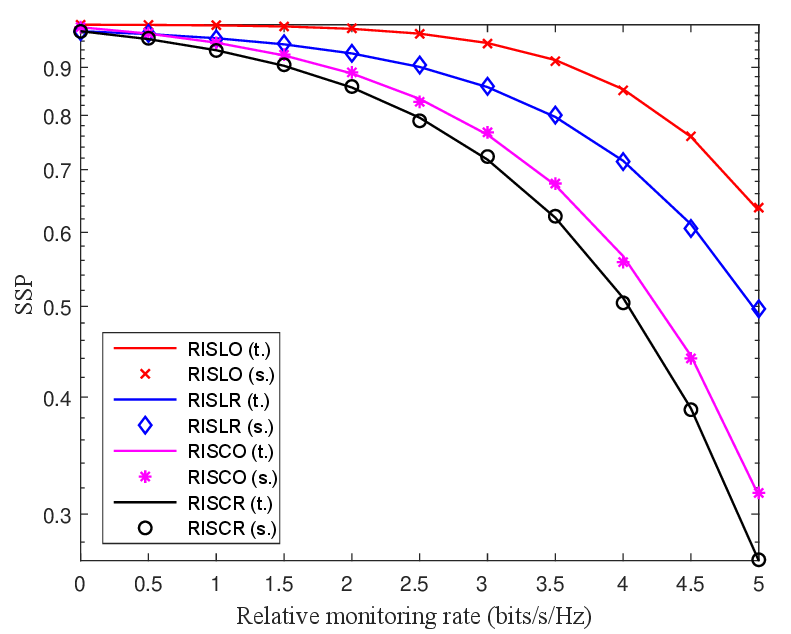}\\
 \caption{Surveillance success probability of RISLR, RISLO, RISCR, and RISCO
schemes versus relative monitoring rate.}
\centering\label{Fig4}} 
\end{figure}

\begin{figure}[h]
\centering {\includegraphics[scale=0.58]{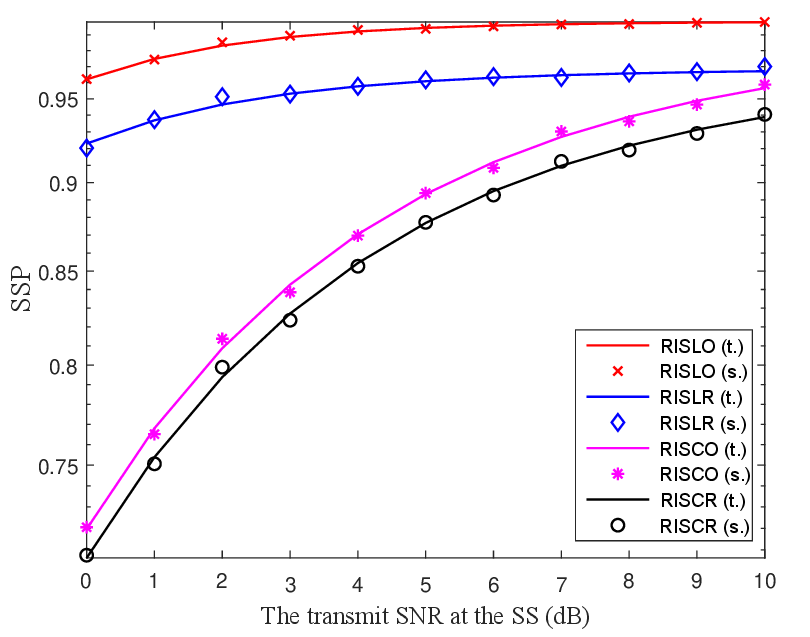}\\
 \caption{Surveillance success probability of RISLR, RISLO, RISCR, and RISCO
schemes versus the transmit SNR at the SS.}
\centering\label{Fig5}} 
\end{figure}

\begin{figure}[tbh]
\centering {\includegraphics[scale=0.55]{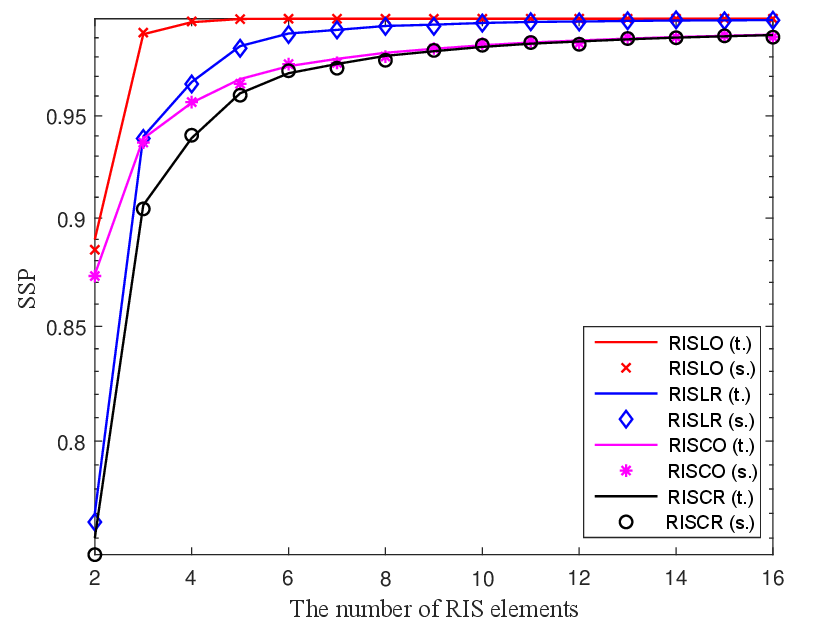}\\
 \caption{Surveillance success probability of RISLR, RISLO, RISCR, and RISCO
schemes versus the number of RIS elements.}
\label{Fig6}} 
\end{figure}

In this section, we present simulation results to evaluate the performance
of the proposed schemes. The numerical values of the system parameters
are listed in Table \ref{tab:para} at the top of the page, unless
otherwise stated. Additionally, for simplicity and without loss of
generality, $\sigma_{n\text{R}}^{2}$ is assumed to be the same for
all $n\in\mathcal{N}$. It is worth mentioning that in Figs. \ref{Fig2}-\ref{Fig6},
theoretical expressions given by \eqref{equa(25)}, \eqref{RISLR},
and \eqref{equa(38)} are represented by solid lines, while Monte-Carlo
simulation results are plotted using dotted markers. The close match
between the theoretical and simulation results confirms the accuracy
of the closed-form analysis.
\begin{table*}[t]
\caption{Simulation parameters}
\label{tab:para}\centering %
\begin{tabular}{c|c|c}
\hline 
Description & Symbol & Value\tabularnewline
\hline 
The variances of reflection channel coefficients & $\sigma_{n\text{R}}^{2},\sigma_{\text{RM}}^{2},\sigma_{\text{SR}}^{2},\sigma_{\text{RD}}^{2}$ & 0.5\tabularnewline
The variances of direct (non-RIS) channel coefficients & $\sigma_{\text{SD}}^{2}$ & 1\tabularnewline
Transmit SNR at the SS & $\gamma_{\text{s}}$ & 10dB\tabularnewline
Jamming SNR & $\gamma_{\text{J}}$ & 10dB\tabularnewline
MSR & $\zeta_{\text{MSR}}$ & 0dB\tabularnewline
The number of jammers & $N$ & 3\tabularnewline
The number of RIS reflecting elements & $L$ & 4\tabularnewline
Relative monitoring rate & $R_{\text{th}}$ & 1bit/s/Hz\tabularnewline
The accuracy versus complexity parameter in the sum approximation & $K$ & 400\tabularnewline
\hline 
\end{tabular}
\end{table*}

Fig. \ref{Fig2} plots the surveillance success probabilities (SSPs)
of RISLR, RISLO, RISCR, and RISCO schemes as a function of jamming
SNR. As the jamming SNR increases, the SSPs of all schemes improve.
In the low jamming SNR region, schemes that incorporate jammer selection
generally achieve higher SSPs, due to the fact that when jammers operate
under power constraints, the channel gains of CJ links become essential,
making jammer selection crucial. This observation is supported by
the analytical results in Section IV. Conversely, in the high jamming
SNR region, RISLO outperforms RISCO in terms of SSPs. As jamming power
increases, both schemes gradually approach their respective performance
limits, consistent with the observations in Remark 2. The significant
gap between these performance limits shows the superior effectiveness
of RISLO over RISCO. The primary reason for this performance gap is
that RISLO leverages CSI from both LS and CJ links for phase shift
design and jammer selection, while RISCO relies only on the CSI of
CJ links. As a result, RISLO makes more efficient use of CSI, leading
to consistently higher SSP performance compared to RISCO.

Fig. \ref{Fig3} depicts the SSPs of RISLR, RISLO, RISCR, and RISCO
schemes versus the monitoring to suspicious ratio (MSR). The low-MSR
region refers to scenarios where suspicious links have higher channel
gains than the monitoring links, and vice versa for the high-MSR region.
Specifically, MSR values below zero means that the monitoring channels
are weaker than the suspicious channels. As can be observed, for the
same jammer selection strategy, the SSP of the RISLO scheme is significantly
higher than that of the RISCO scheme. Again, schemes that employ optimal
jammer selection achieve higher SSPs compared to those without selection.
However, this improvement comes at the cost of acquiring additional
CSI knowledge, especially in the low MSR region. 

Fig. \ref{Fig4} plots the SSPs of RISLR, RISLO, RISCR, and RISCO
schemes versus RMR, which represents the target reliability of the
monitoring process. The results indicate that schemes employing OJS
consistently achieve significantly better performance than conventional
RIS schemes. Moreover, as the target RMR increases, the performance
gap between RISLO and RISCO becomes more pronounced. The results in
Fig. \ref{Fig4} again highlights the superiority of RISLO, which
benefits from a more comprehensive CSI database.

Fig. \ref{Fig5} illustrates the SSPs of RISLR, RISLO, RISCR, and
RISCO schemes as a function of the transmit SNR at the SS. Although
the SSPs of all schemes improves as the transmit power of suspicious
communication increases, this may not happen when the illegal party
attempts to make covert communication quietly. Typically, if the illegal
party wants the message only to be known to suspicious nodes, they
will limit transmit power to avoid being detected. Consequently, in
the low-SNR regime at the SS, schemes incorporating jammer selection
achieve significantly higher SSPs compared to those without jammer
selection.

Fig. \ref{Fig6} plots the SSPs of RISLR, RISLO, RISCR, and RISCO
schemes against the number of RIS elements in the suspicious link.
As the number of RIS elements increases, the SSPs of all schemes improve
significantly. In scenarios with a small number of RIS elements, schemes
that employ jammer selection outperform those without it. However,
as the number of RIS elements increases, RISLO and RISCO gradually
converge to distinct performance limits, as explained in Remark 2.
This divergence underscores the performance advantage of RISLO over
RISCO, which stems from its broader CSI database for phase shift optimization
and jammer selection. These findings further reinforce that that CSI
acquisition and phase shift accuracy are fundamental to the effectiveness
of RIS-aided communications.

\section{Conclusion}

In this paper, we presented a RIS-aided wireless surveillance system
assisted by multiple jammers. We then proposed RISLO and RISCO schemes,
each incorporating different jammer selection strategies. We derived
the SSP expressions that reveal a tradeoff between monitoring performance
and jammer implementation complexity, which underscores the critical
role of CSI utilization. Simulation results not only confirmed our
closed-form analysis, but also demonstrated the advantage of the proposed
jammer selection strategies in enhancing surveillance performance.

\appendix{\label{appendix}}

We first establish that the phase shifts of RIS, which maximize the
average gain of LS link, appear random from the perspective of the
CJ link. Specifically, this ``randomness" corresponds to a uniform
distribution on the range $[0,2\pi)$. From \eqref{equa(4)}, the
phase shifts are determined from the phase angles of ${h_{\text{R}_{l}\text{M}}}$
and ${h_{\text{SR}_{l}}}$ for $l\in\mathcal{L}$ representing $L$
reflecting links. Given that these channel coefficients follows a
complex Gaussian distribution, the resulting phase shifts exhibit
a uniform distribution over $[0,2\pi)$. Moreover, due to the statistical
independence between suspicious links and monitoring links, the phase
shifts are completely random for other links \cite{random_phase_covert}. 

Let us consider the RISLO scheme first. Take the CJ channel analysis
as an example, and the next step is to derive the distribution of
the cascaded channel gain $Q_{n}^{\text{RISLO}}=|{\bf {h}}_{\text{RD}}^{\text{H}}\boldsymbol{\Theta}^{\text{RISLO}}{\bf {h}}_{n\mathit{\textrm{R}}}|^{2}$
with random phase shifts. We can explicitly rewrite the cascaded channel
as 
\begin{equation}
\begin{aligned}{\bf {h}}_{\text{RD}}^{\text{H}}\boldsymbol{\Theta}^{\text{RISLO}}{\bf {h}}_{n\mathit{\textrm{R}}}= & \underbrace{\sum_{l=1}^{L}\left|h_{\text{R}_{l}\text{D}}\right|\left|h_{\text{\ensuremath{{\displaystyle {\scriptstyle n}}}R}_{l}}\right|\cos\left(\phi_{\text{R}_{l}\text{D}}+\phi_{\text{\ensuremath{{\displaystyle {\scriptstyle n}}}R}_{l}}+{\phi_{l}}\right)}_{X_{1}}\\
 & +j\underbrace{\sum_{l=1}^{L}\left|h_{\text{R}_{l}\text{D}}\right|\left|h_{\text{\ensuremath{{\scriptstyle n}}R}_{l}}\right|\sin\left(\phi_{\text{R}_{l}\text{D}}+\phi_{\text{\ensuremath{{\displaystyle {\scriptstyle n}}}R}_{l}}+{\phi_{l}}\right)}_{X_{2}},
\end{aligned}
\label{A1}
\end{equation}
where $l\in\mathcal{L}$, and $j$ is the imaginary unit. In \eqref{A1},
${\phi_{l}}$, given by\eqref{equa(4)}, is uniformly-distributed
and independent of $\phi_{\text{R}_{l}\text{D}}$ and $\phi_{\text{\ensuremath{{\displaystyle {\scriptstyle n}}}R}_{l}}$.
Using the periodicity of trigonometric functions $\cos$ and $\sin$,
for $\varphi\in[0,2\pi)$ randomly, it follows that $\text{E}(\cos\varphi)=\text{E}(\sin\varphi)=0$
and $\text{E}(\cos^{2}\varphi)=\text{E}(\sin^{2}\varphi)=\frac{1}{2}$.
Applying the central limit theorem for a large number of reflecting
elements \cite{random_rotation}, we deduce that $X_{1}\sim\mathcal{CN}(0,\frac{\sigma_{\text{SD}}^{2}}{2}+\frac{L\sigma_{\text{RD}}^{2}\sigma_{\text{SR}}^{2}}{2}),X_{2}\sim\mathcal{CN}(0,\frac{\sigma_{\text{SD}}^{2}}{2}+\frac{L\sigma_{\text{RD}}^{2}\sigma_{\text{SR}}^{2}}{2})$.
Recalling \eqref{A1}, we know ${\bf {h}}_{\text{RD}}^{\text{H}}\boldsymbol{\Theta}^{\text{RISLO}}{\bf {h}}_{n\mathit{\textrm{R}}}=X_{1}+jX_{2}$,
 and thus $Q_{n}^{\text{RISLO}}$ follows a Rayleigh distribution,
given by
\begin{equation}
F_{Q_{n}^{\text{RISLO}}}(y)=1-{e^{-\frac{y}{L\sigma_{n_{\text{R}}}^{2}\sigma_{_{\text{RD}}}^{2}}}}.\label{equa(14)-1}
\end{equation}

\begin{figure}
\centering{\includegraphics[scale=0.55]{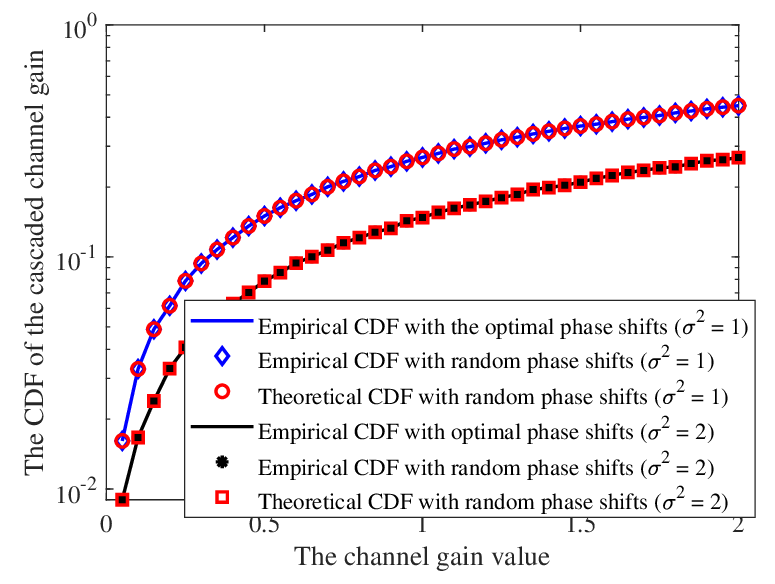}\caption{ Illustration of the CDF of $Q_{n}^{\text{RISLO}}$ for different
phase shift configurations. Notably, the CDF of $Q_{n}^{\text{RISLO}}$
with optimal phase shift given by \eqref{equa(3)} shows the system
property of the proposed scheme with partial CSI, and the other two
validate the correctness of our analytical formulation.}
\label{Fig7}}
\end{figure}

To further support the above analytical formulation, we present numerical
results to verify its accuracy. Specifically, Fig. \ref{Fig7} plots
the theoretical CDF of $Q_{n}^{\text{RISLO}}$ with the random RIS
phase shifts, i.e., given in \eqref{equa(14)-1}, the empirical CDF
with the optimal phase shifts using \eqref{equa(3)}, and the empirical
CDF with random phase shits.  As can be seen clearly, the three CDF
curves are indeed the same, which means that the phase shifts that
maximize the average gain of SS-LM transmission appear random from
the perspective of the LM-SD transmission. 

\end{document}